\documentclass[conference]{IEEEtran}
\pdfoutput=1

\usepackage{graphicx}  

\usepackage{amsmath}   
\begin{document}

\title{Comparison of reconstruction algorithms for digital breast
tomosynthesis}

\author{\authorblockN{I. Reiser, J. Bian, R. M. Nishikawa,
    E. Y. Sidky, and X. Pan}
\authorblockA{Department of Radiology\\
The University of Chicago\\
Chicago, IL 60637\\
}}

\maketitle

\begin{abstract}
Digital breast tomosynthesis (DBT) is an emerging modality for breast
imaging. A typical tomosynthesis image is reconstructed from
projection data acquired at a limited number of views over a limited
angular range. In general, the quantitative accuracy of the image can  
be significantly compromised by severe artifacts and non-isotropic
resolution resulting from the incomplete data. Nevertheless, it has
been demonstrated that DBT may yield useful information for 
detection/classification tasks and thus is considered a
promising breast imaging modality currently undergoing 
pre-clinical evaluation trials. 
%
%
The purpose of this work is to conduct a preliminary, but systematic,
investigation and evaluation of the properties of
reconstruction algorithms that have been proposed for DBT. We 
use a breast phantom designed for DBT evaluation
to generate analytic projection data for a typical DBT configuration,
which is currently undergoing  pre-clinical evaluation.      
The reconstruction algorithms under comparison 
include (i) filtered backprojection (FBP), (ii) expectation maximization
(EM), and (iii) TV-minimization algorithms. 
Results of our studies indicate that
FBP reconstructed images are generally
noisier and demonstrate lower in-depth resolution than 
those obtained through iterative reconstruction and that
the TV-minimization reconstruction
yield images with reduced artifacts as compared to that obtained
with other algorithms under study.   
\end{abstract}


%
\IEEEpeerreviewmaketitle

\section{Introduction}
The breast cancer death rate declined  by an average of 2.3\% per
year between 1990 and 2002 \cite{CancerFacts:2006}. This has been
attributed to earlier detection of breast cancer through mammography
screening. In a mammographic exam, two x-ray projections of each 
breast are acquired in cranio-caudal and mediolateral-oblique views.  
While mammography is considered the gold standard in breast imaging,
mass lesion detection is limited by anatomic background structures
\cite{Burgess:2001}, which can obscure lesions or mimic a lesion
appearance.  

This limitation can be significantly reduced in tomosynthesis imaging,
in which a volume image is reconstructed from projection data acquired
at a limited number of views over a limited angular range. Typically,
the number of views ranges between 11 and 21, whereas the angle ranges
between 15 and 50 degrees. Tissue structures that are overlaying in
conventional mammography can be resolved in the reconstructed volume, 
and lesions can thus become more conspicuous.  
Because of the data incompleteness in DBT
the quantitative accuracy of DBT images is generally 
compromised significantly by severe artifacts and 
non-isotropic resolution. However, the purpose of DBT is not to
provide an accurate attenuation map of the breast. Instead, DBT is
aiming only at providing a clinically useful image, i.e., to
``remove visual clutter'' \cite{Dobbins:2003}, in terms of
detection/classification tasks.  

The advent of digital detectors for mammography has made DBT become
tangible since the late 1990s. The development of DBT was initialized
by a breast mammographer, Daniel Kopans and colleagues, at the
Massachusetts General Hospital \cite{Niklason:1997}. Currently, a
number of DBT prototypes have been built, implementing a range of
acquisition geometries
\cite{Ren:2005,Mertelmeier:2006,Maidment:2006,Zhang:2006}. In general,
reconstruction algorithms for DBT that have been  explored fall into
two broad categories, namely filtered backprojection
\cite{Ren:2005,Mertelmeier:2006}, and iterative methods
\cite{Zhang:2006, Wu:2003}.   

Accurate image reconstruction in DBT is challenging because 
of the high degree of data incompleteness. Therefore no
analytic solution exists for quantitatively accurate
image reconstruction from
projection data. On the other hand, iterative methods have been used
to address image reconstruction in DBT, and they appear to yield
visually improved images over those obtained with the non-iterative
algorithms. In DBT, high in-plane resolution is required in order to
visualize microcalcification clusters. The in-depth resolution of
the images 
is poor largely because of the limited angular range. Typically, the
image sampling is non-isotropic, with an in-plane resolution of
up to a factor of 10 higher than the in-depth resolution, which
can lead to artifacts during reconstruction. While this anisotropy
could in principle be remedied by increasing the sampling grid
resolution, this approach is not practically feasible 
because of the large data size.  
Currently, a typical tomosynthesis image that adequately covers
the breast volume consists of about 1.32$^8$ voxels 
(corresponding to a resolution of $0.1\times 0.1 \times1$~mm$^3$).

The purpose of this work is to compare reconstruction methods that are
currently used for and/or have been proposed for DBT reconstruction. 
Our comparison study is based on an analytic phantom that we have
developed for the purpose of DBT algorithm evaluation.
Dominant features of clinical relevance
have been incorporated into this model, namely the
overall breast shapes which causes a drop-off in intensity towards the
skinline, spherical structures to represent tumor lesions, as well as
microcalcification clusters.

\section{Breast Phantom}
\begin{figure}
\centering
\includegraphics[width=1.0in]{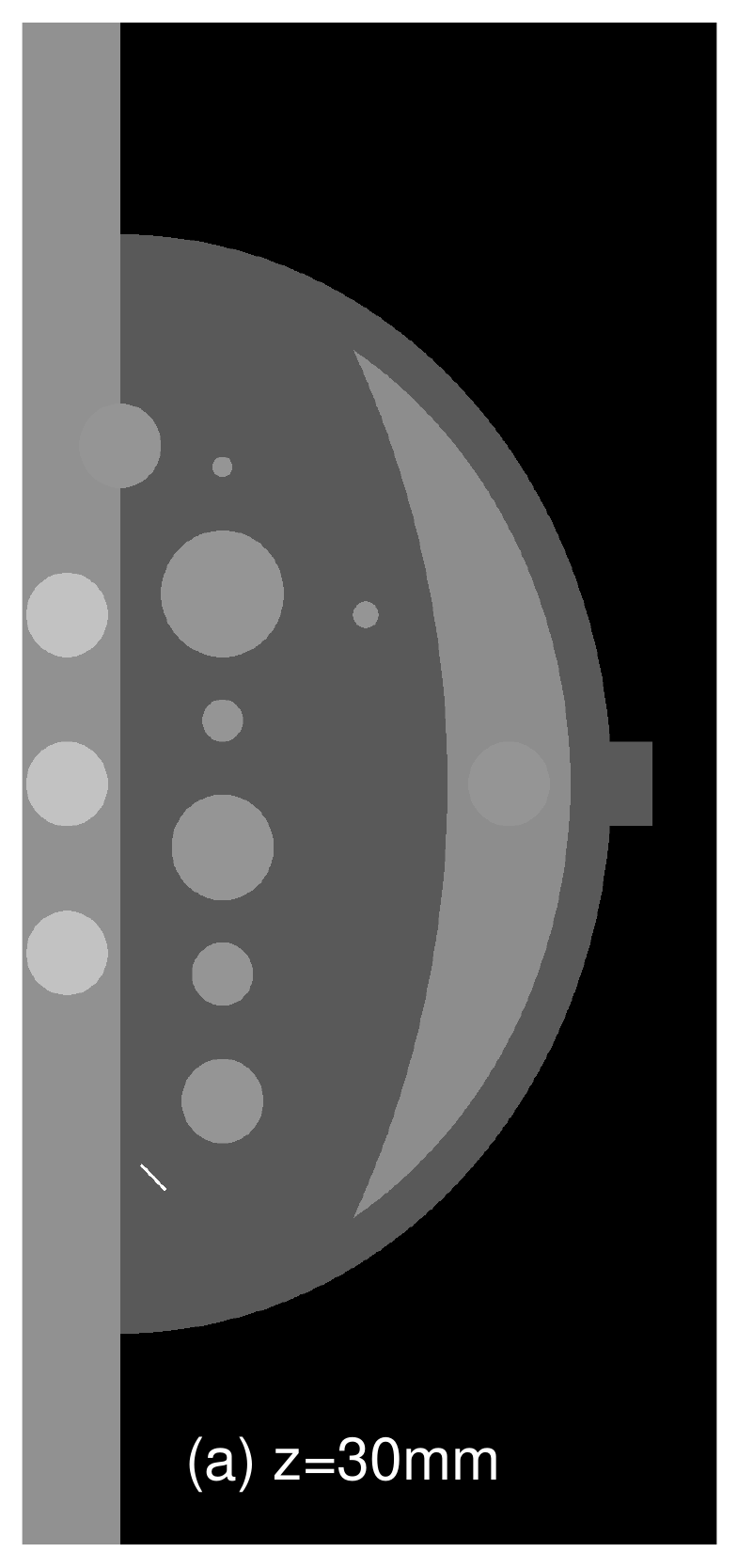}
\includegraphics[width=0.7in]{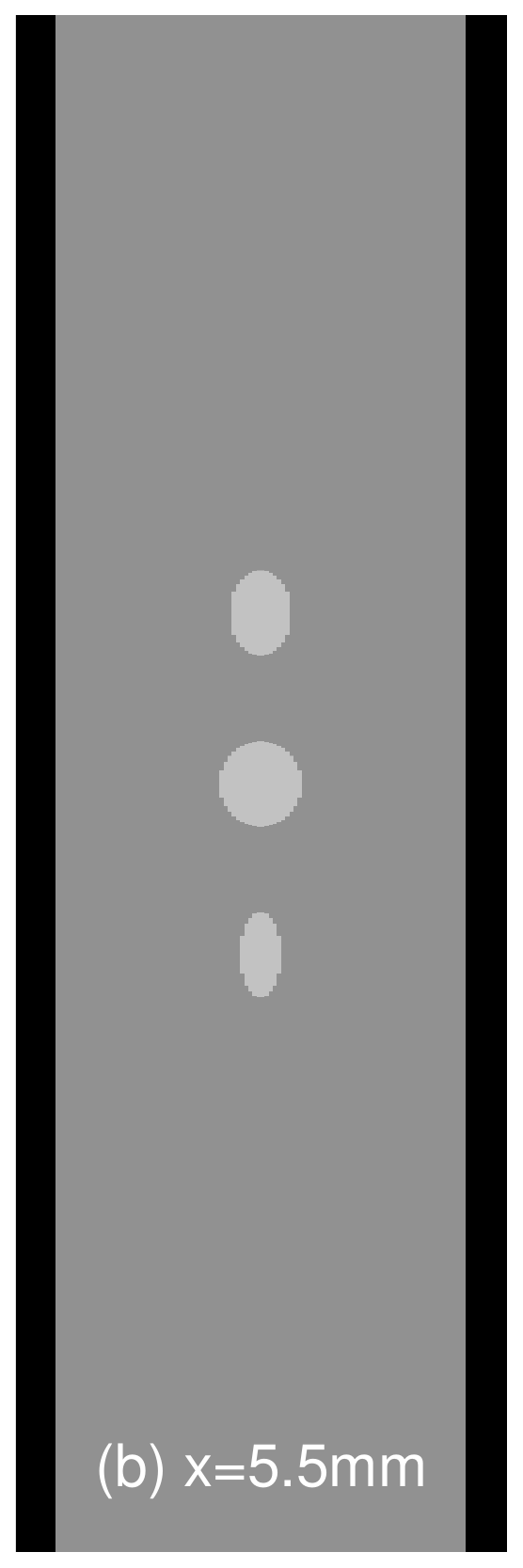}
\includegraphics[width=0.7in]{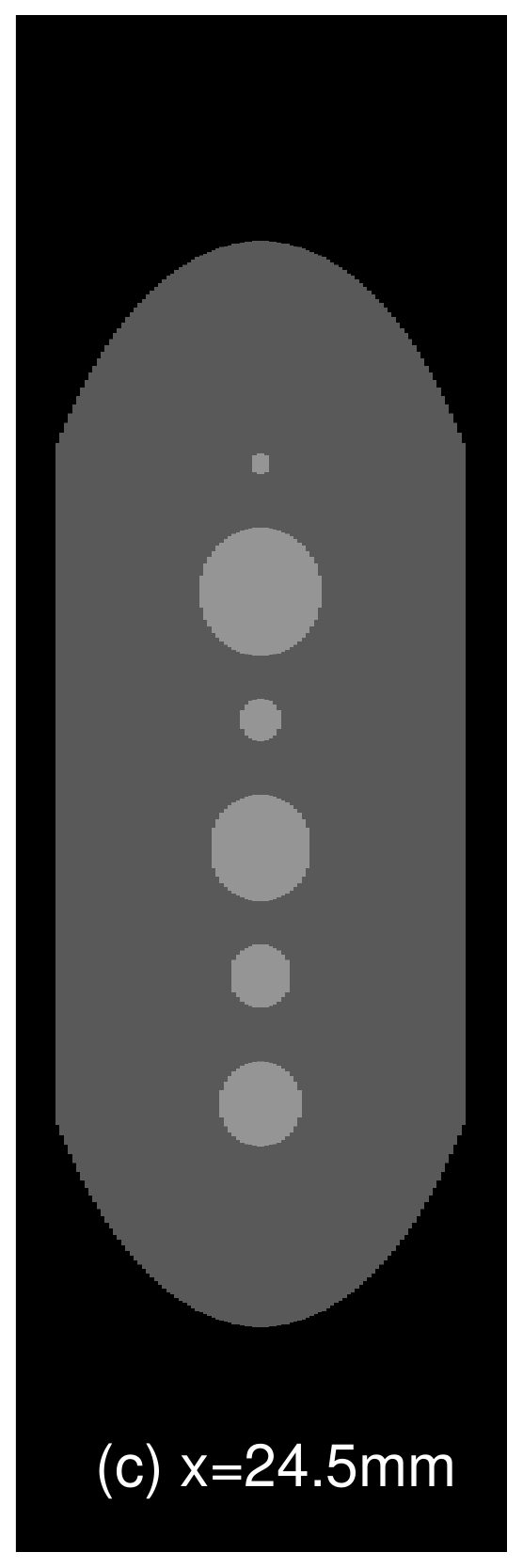}
\includegraphics[width=0.7in]{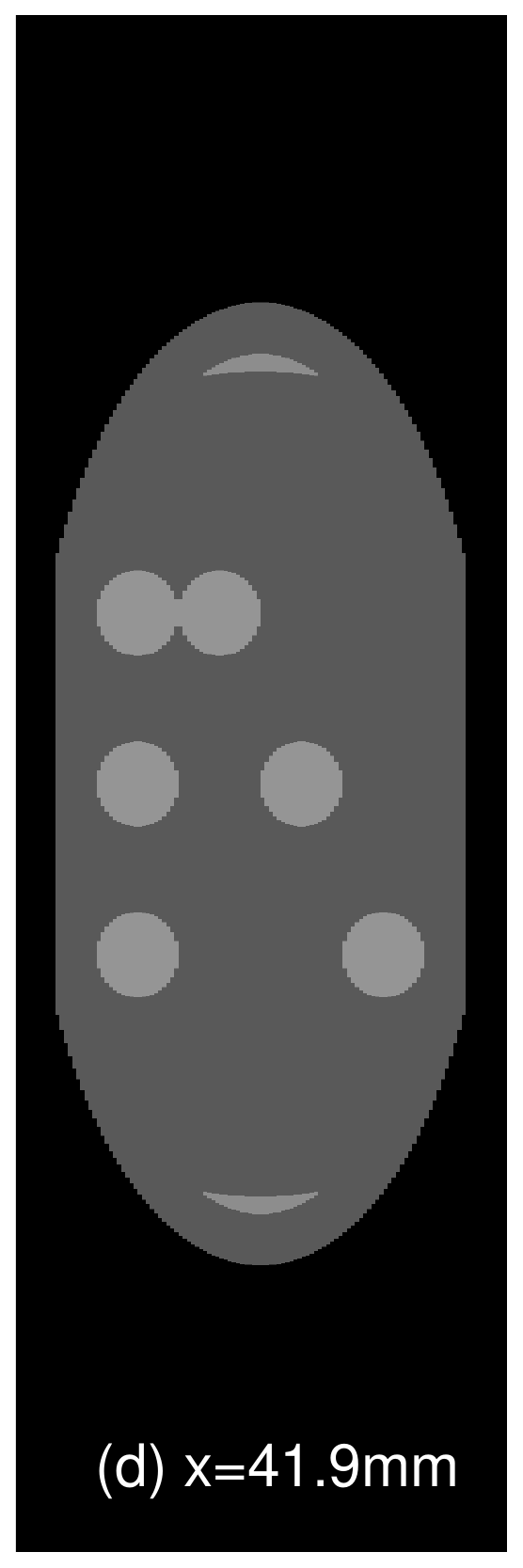}
\caption{
The breast phantom used as the basis for comparison of 
reconstruction algorithms in DBT. Images of the breast phantom
within 2D planes specified by (a) $z=30$~mm, by   
(b) $x=5.5$~mm,  (c) $x=24.5$~mm,  and  (d) $x=41.9$~mm, 
respectively. }
\label{fig:phantom}
\end{figure}
A breast phantom was designed to provide a basis for algorithm
comparison. It consists of objects with different geometric shapes to
allow for analytic computation of  the projection data. 
The overall shape of the compressed breast of 50~mm thickness
is modeled by a truncated ellipsoid.  
Attenuation coefficients at a photon energy of 30 keV 
\cite{Johns:1987} were used for   
phantom materials simulating the fibroglandular, tumour
and muscle tissues, as well as microcalcifications in the
breast.  In Fig.~\ref{fig:phantom}, we display the phantom images
within 2D planes specified by (a) $z=30$~mm, by   
(b) $x=5.5$~mm,  (c) $x=24.5$~mm,  and  (d) $x=41.9$~mm, 
respectively.  As shown in Fig.~\ref{fig:phantom}a, 
the crescent-shaped region, representing the fibroglandular 
tissue region of the breast, is attached to a rectangular slab 
of higher attenuation, simulating the pectoralis muscle. 
It can be observed in Fig.~\ref{fig:phantom} that  
numerous test objects of different sizes and contrast 
levels are embedded in the phantom for 
simulating mass lesions and microcalcifications. In particular, 
a row of three ellipsoids is embedded in the pectoralis 
muscle (see Fig.~\ref{fig:phantom}(b)), with
equal in-plane diameters but varying flatness, thus allowing one to 
evaluate whether shapes of equal in-plane profile, but different 
in-depth profiles, can be resolved.  Furthermore, 
six spheres of different diameters, ranging 
from 5~mm to 15~mm, are equally distributed in the fatty 
area of the breast (see Fig.~\ref{fig:phantom}(c)). 
It can also be observed in Fig.~\ref{fig:phantom}(d) that
three pairs of stacked spheres of identical diameter
$d=10$~mm are embedded in the breast region with different 
spacings of $d$, $2d$, and $3d$, respectively, for these  
pairs. Finally, two clusters of small spheres of diameters
0.3~mm and 0.15~mm are included to model microcalcifications. 

\section{Scanning Configuration}
In this work, we consider a tomosynthesis scanning configuration
with parameters similar to those of the first GE tomosynthesis
prototype for breast imaging \cite{Wu:2003}. 
In this configuration, the source is acquiring data at 11 projection
views uniformly distributed over an arc of 90 degrees.
The source-to-detector distance is 660 mm, whereas the distance
between the source and center-of-rotation is 460 mm. The detector
plane is perpendicular to the line connecting 
the x-ray source and the detector center, 
at the center (6th) projection, 
and the detector remains stationary while the x-ray source
rotates along the arc. 
The detector size is 180~mm$\times$85~mm with the shorter side
perpendicular to the plane of rotation, and the detector-bin size is
200~microns. In our configuration, the object center is located 
50~mm above  the detector. 

\section{Reconstruction Algorithms}
Noiseless projection data were computed analytically from the breast
phantom for the scanning configuration described above. 
Noisy data were created by adding Poisson noise to the noiseless
projection data. The photon counts in projection data were weighted 
to account for variations in the source-to-detector-bin distance 
as well as variations in the effective area of the detector-bin 
surface due to oblique x-ray incidence. We have used a typical
clinical exposure to determine the added noise level. 
From the projection data, images within 3D arrays with a voxel
size of  0.2 x 0.2 x 1mm$^3$ were reconstructed by use of 
different reconstruction algorithms.

\subsection{The FBP algorithm}
For the filtering and backprojection reconstruction (FBP) \cite{Wu:2004},
each projection image is filtered with a ramp filter and then
back-projected using a cone-beam geometry. A von Hann window is applied
to supress high frequencies. 

\subsection{The EM algorithm}
In each iteration of the expectation-maximization
algorihtm, the current image estimate 
$\vec{f}_{k+1}$ is updated by 
\begin{equation}
\vec{f}_{k+1} = \vec{f}_{k}\, \frac{1}{s}\,M^+\{\frac{\vec{g}}{M \vec{f}_{k}}\}
\end{equation}
where $\vec{g}$ is the projection data, $M$ and $M^+$ 
are the forward and backprojection operators, $s=\sum M$, and
$\frac{\vec{g}}{M \vec{f}_{k}}$ indicates a division between
the vector elements.
In our implementation, a matched projector/backprojector pair is used
with Gaussian smoothing between iterations. 

\subsection{The TV-minimization algorithm}
In the TV-minimization algorithm \cite{Sidky:2006}, 
the image is obtained by solving an optimization problem:
\begin{equation}
\label{realopt}
\vec{f}^* = \mbox{argmin} \| \vec{f} \|_{TV} \mbox{  such that  }
\left| M \vec{f} - \vec{g} \right| \leq \epsilon ,
\end{equation}
where $\vec{f}$ and $\vec{g}$ are discrete image and data, and M is
the linear operator representing the cone-beam forward
projection and $\vec{f}^*$ is the reconstructed image. The parameter
$\epsilon$ can be selected for controlling the impact level of 
potential data inconsistency on the image reconstruction.   

\section{Results}

\begin{figure}
\centering
\includegraphics[width=1.0in]{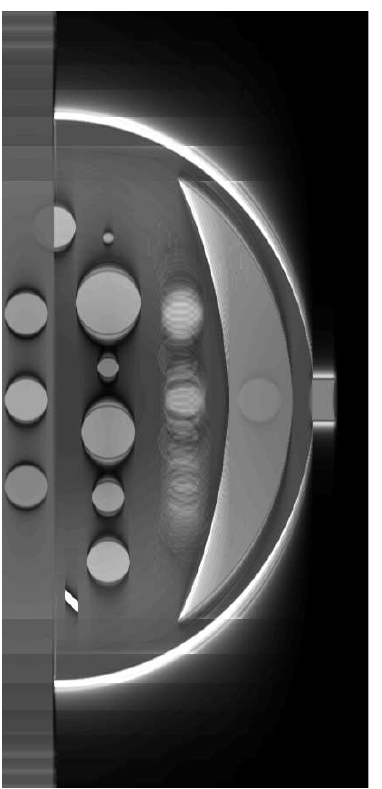}
\includegraphics[width=1.0in]{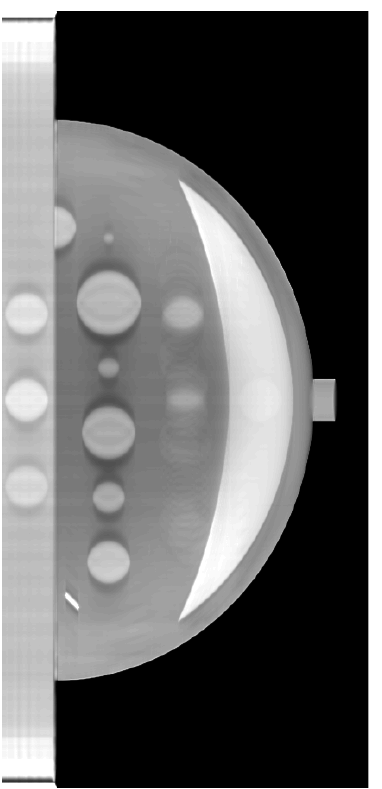}
\includegraphics[width=1.0in]{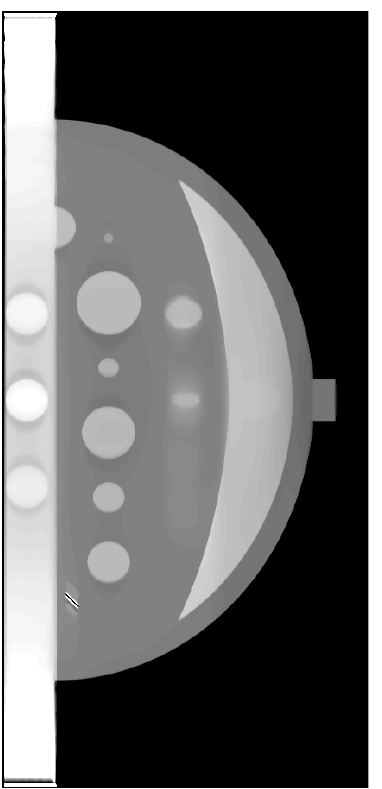}\\
\includegraphics[width=1.0in]{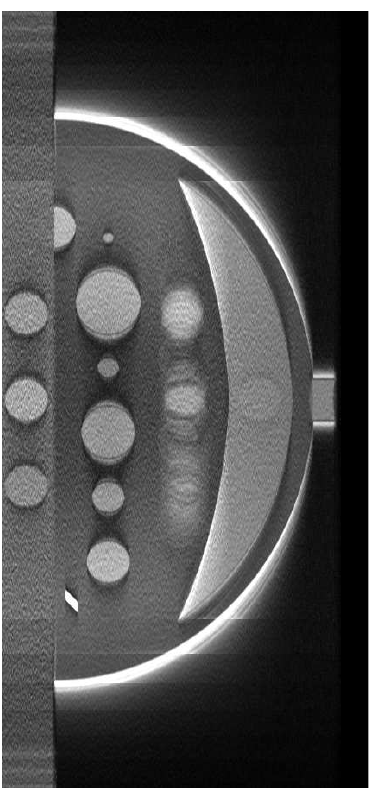}
\includegraphics[width=1.0in]{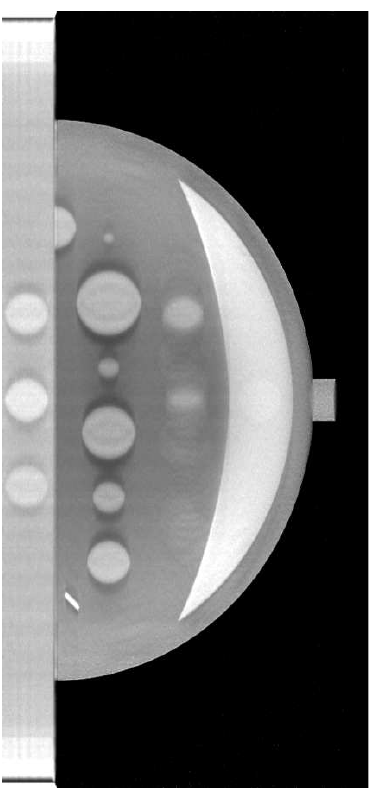}
\includegraphics[width=1.0in]{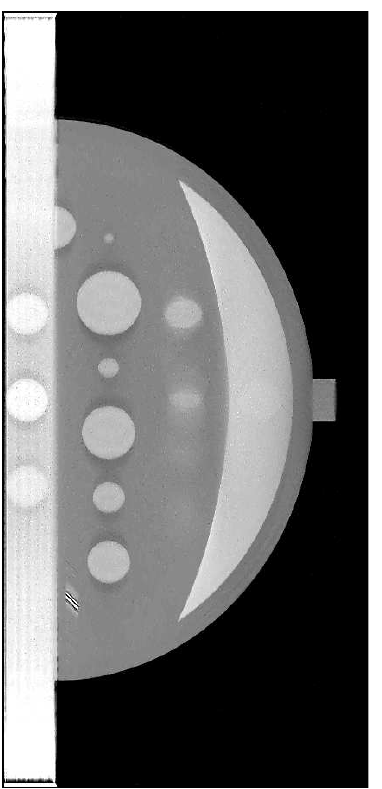}
\caption{Images reconstructed within a 2D slice specified 
by $z=30$~mm , reconstructed from noiseless data (top row) and 
from noisy data (bottom row) by use of the FBP (left column)
EM (middle column), and TV (right column) algorithms,
respectively. }
\label{fig:CUTXY}
\end{figure}

Using the tomosynthesis data produced from the breast phantom
with the scanning configuration described,  
we have carried out a preliminary numerical studies to compare the 
performance of the FBP, EM, and TV algorithms. In 
Fig.~\ref{fig:CUTXY}, we displayed the images within
the plane $z=30$~mm reconstructed by use of the FBP (left),
EM (middle), and TV (right) algorithms. The true image within
the corresponding slice is shown in Fig.~\ref{fig:phantom}(a). 
As expected, because of the strong data incompleteness in
tomosynthesis, all reconstructions exhibit obvious
artifacts.  Out-of-plane objects create 
conspicuous ghosting artifacts in all reconstructions, 
but most severe in the FBP reconstruction. 
We refer to these artifacts as structure noise. 
Among these algorithms, TV algorithm appears to 
produce more uniform images than do others. 
Furthermore, in the FBP reconstruction, 
edges are "enhanced" by the ramp filtering. The noisy 
reconstructions are visually similar, suggesting that the image
noise is dominated by structure noise caused by out-of-plane
objects. It is also observable that the noise properties of 
each algorithm is different. The EM reconstruction appears 
most smooth, while the TV reconstruction exhibits some ``speckle
noise'', i.e., spikes, in an otherwise smooth image.  


\begin{figure}
\centering
\includegraphics[width=3in]{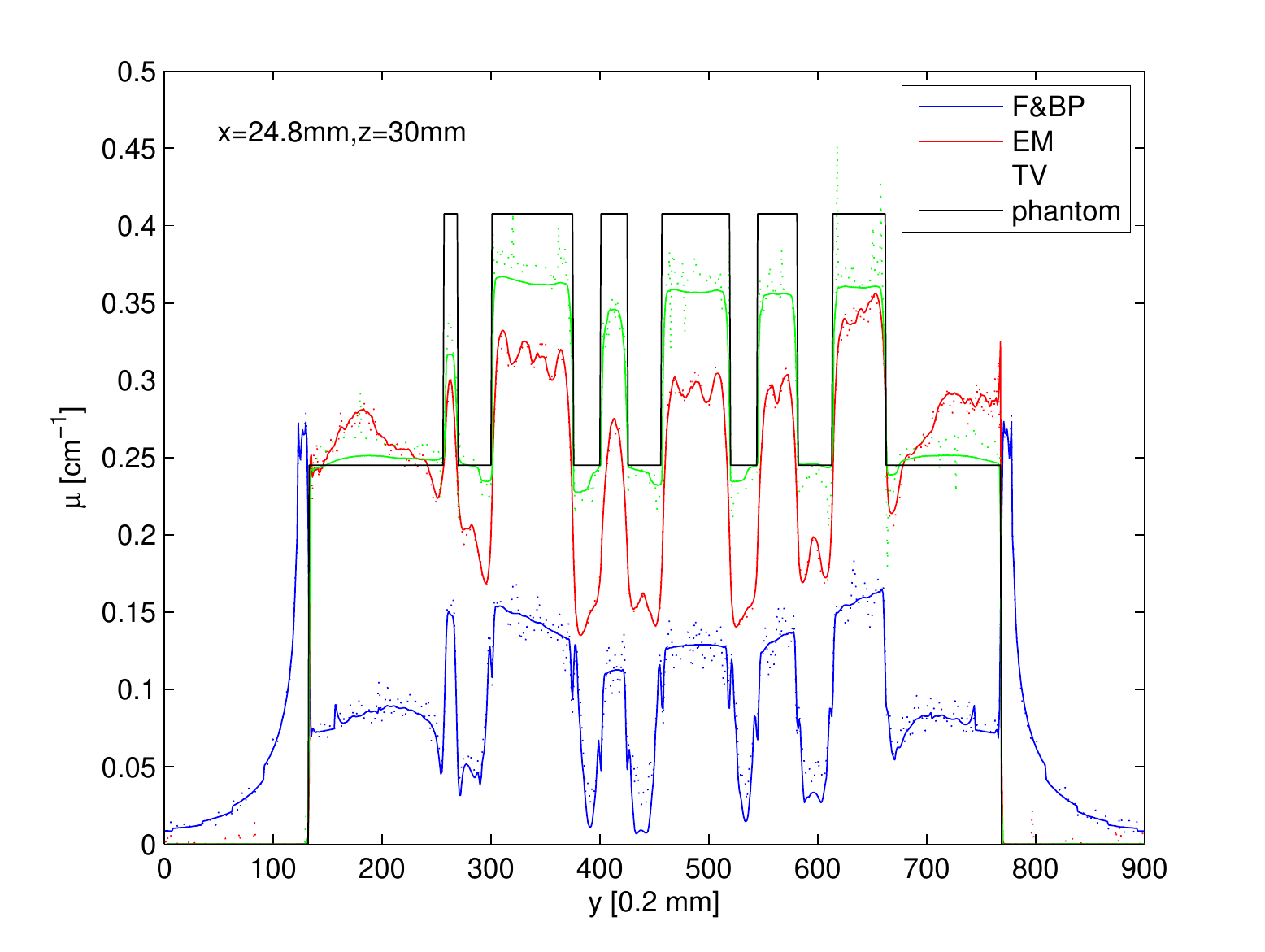}
\caption{Profiles on the line specified by $x=24.8$ mm and $z=30$~mm
  in images displayed in Fig.~{\ref{fig:CUTXY}} obtained with the FBP (blue),
EM (red), and TV (green) algorithms, respectively. Solid curves are
for noiseless images, dotted curves for noisy images. For comparison,
the true profile is depicted by the black solid curve.}
\label{fig:LineProfX248}
\end{figure}

\begin{figure}
\centering
\includegraphics[width=3in]{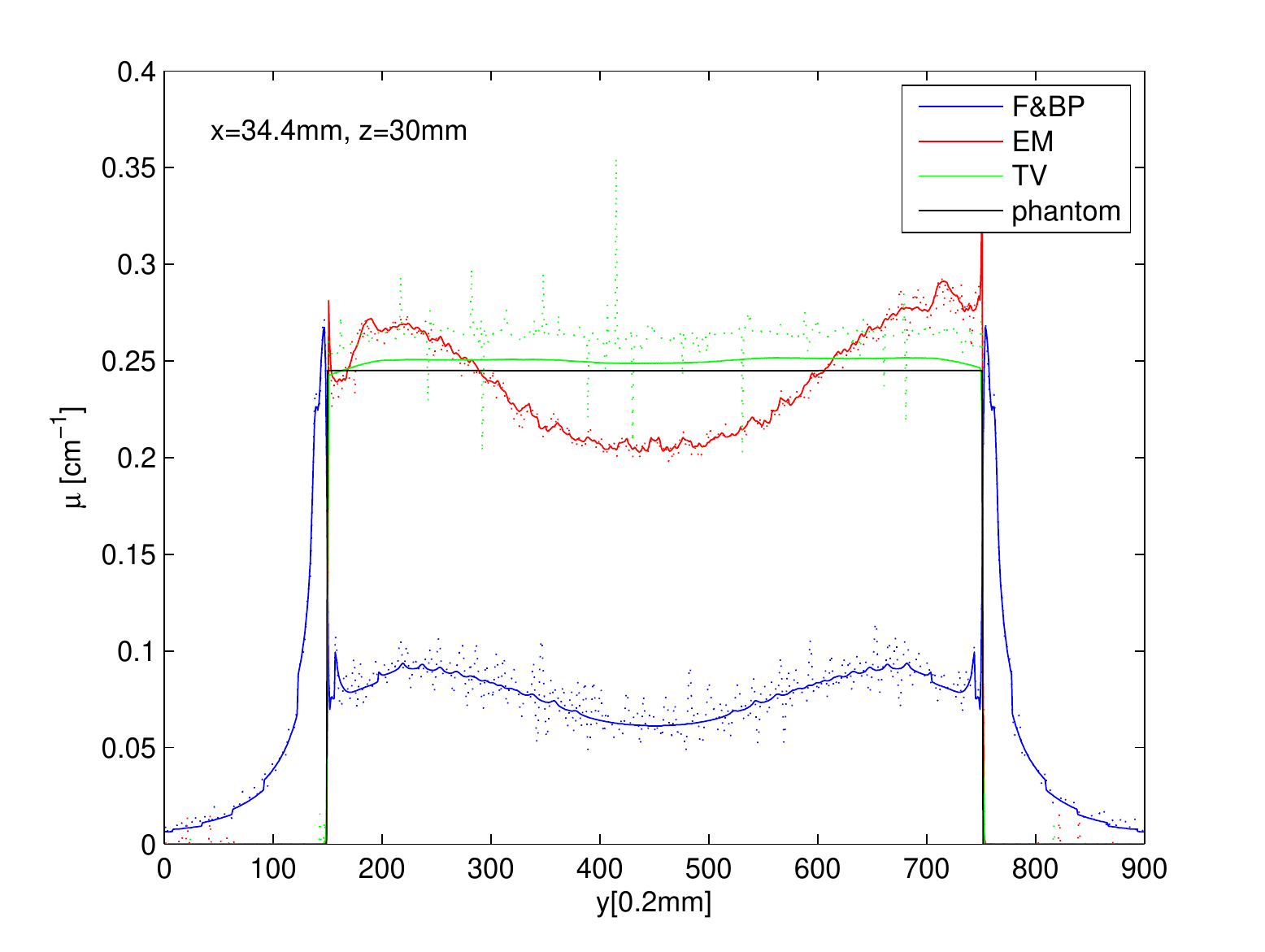}
\caption{Profiles on the line specified by $x=34.4$~mm. 
  Reconstruction algorithms and line types are 
identical to those in Fig.~\ref{fig:LineProfX248}.}
\label{fig:LineProfX344}
\end{figure}

In Figs.~\ref{fig:LineProfX248}, and
\ref{fig:LineProfX344}, we show the image profiles on three different
lines  through the reconstructed image within the center plane (i.e.,
$z=30$~mm) of the image. The reconstruction properties discussed above
can be observed in these profile results. Clearly, the FBP
reconstruction appears most noisy, which is expected because the ramp
filtering tends to amplify noise.  
There is a considerable DC shift for FBP reconstruction, as
well as for the EM reconstruction. 
From the profile across spheres of varying diameters
in Fig.~\ref{fig:LineProfX248}, it can be observed that
that in-plane resolution is maintained
in DBT, which is the reason for the clinical usefulness of
DBT. The profile in Fig.~\ref{fig:LineProfX344} is on a line
through a uniform section of the phantom. Uniformity appears 
to be best reproduced by the TV reconstruction. 
In the TV reconstruction, a shift between
noiseless and noisy data can be observed.

\begin{figure}
\centering
\includegraphics[width=1.1in]{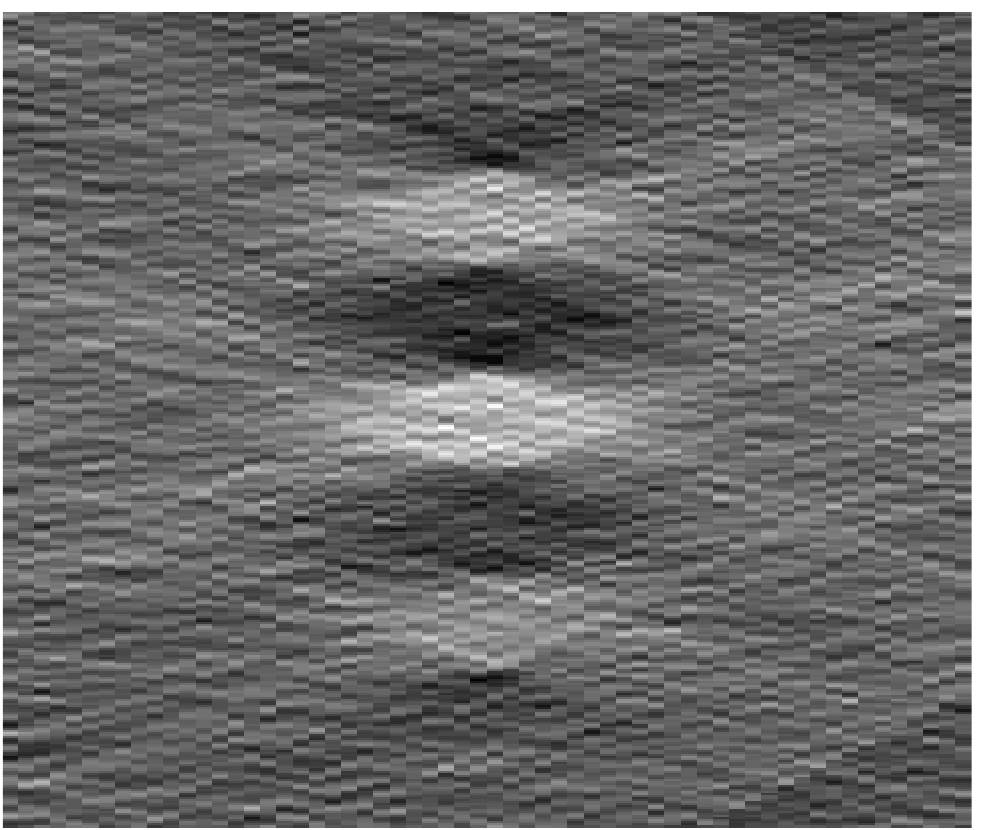}
\includegraphics[width=1.1in]{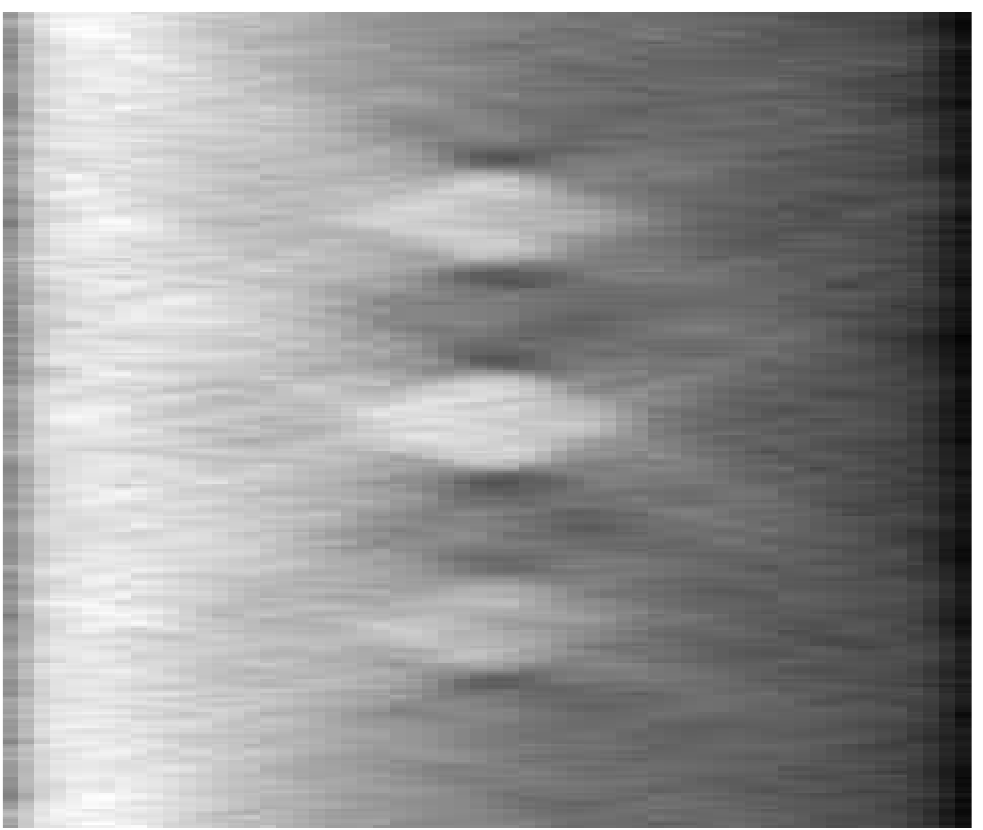}
\includegraphics[width=1.1in]{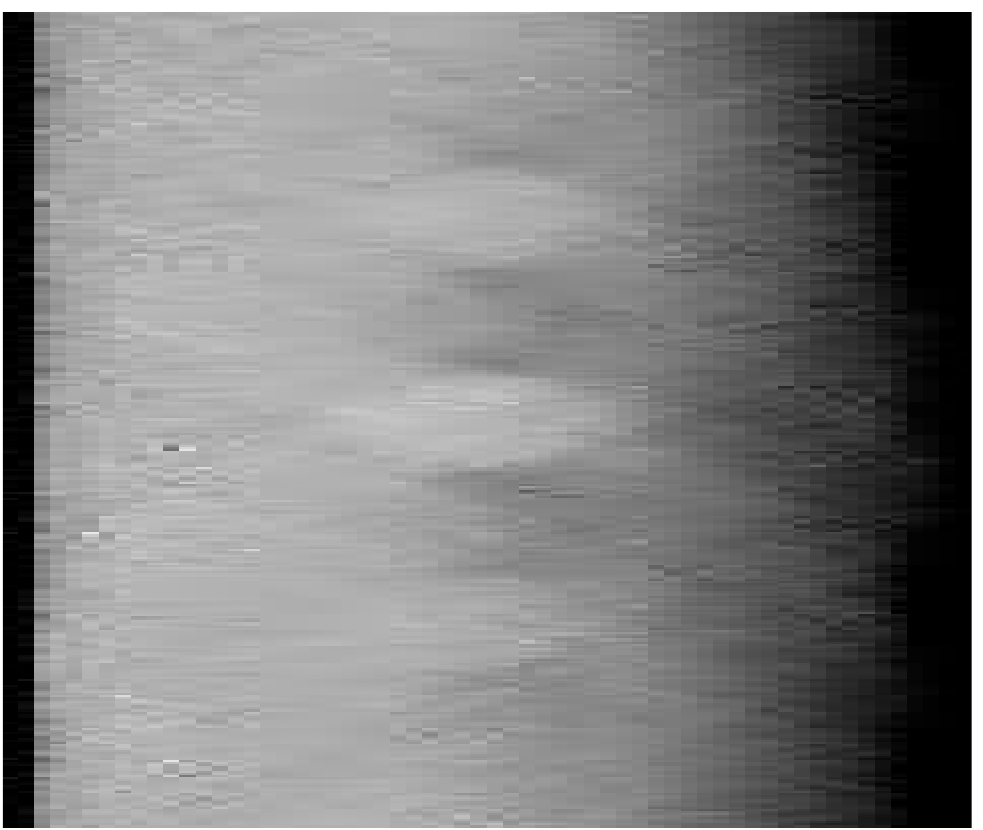}\\
\includegraphics[width=1.1in]{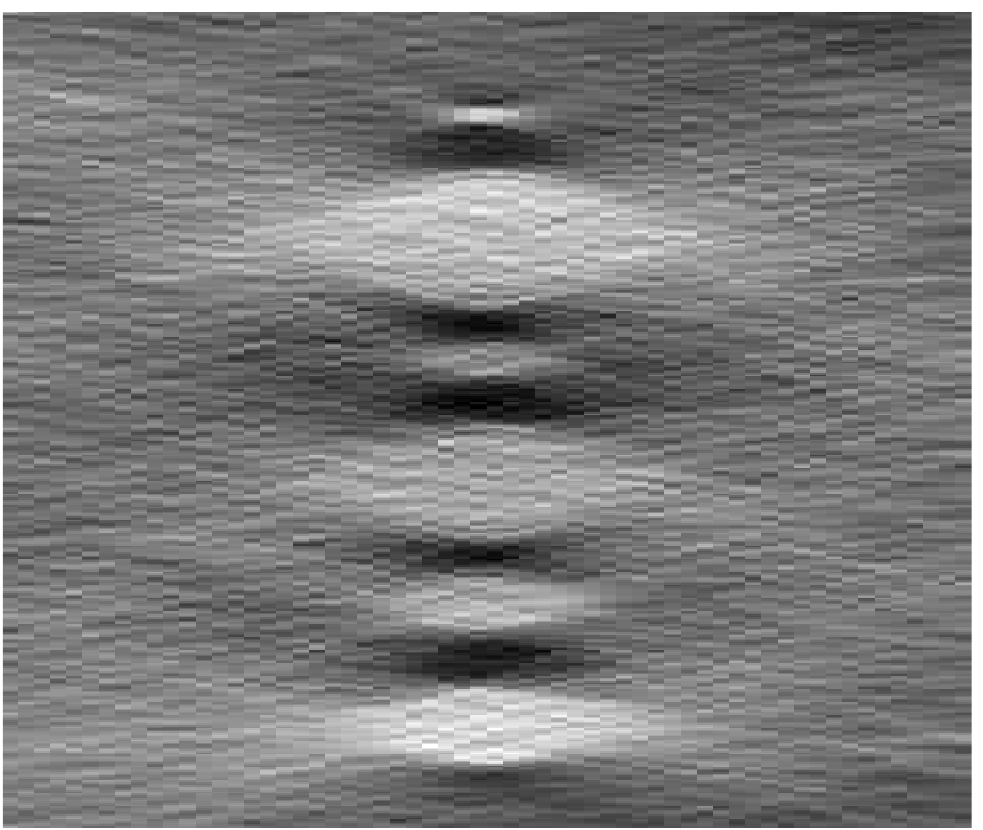}
\includegraphics[width=1.1in]{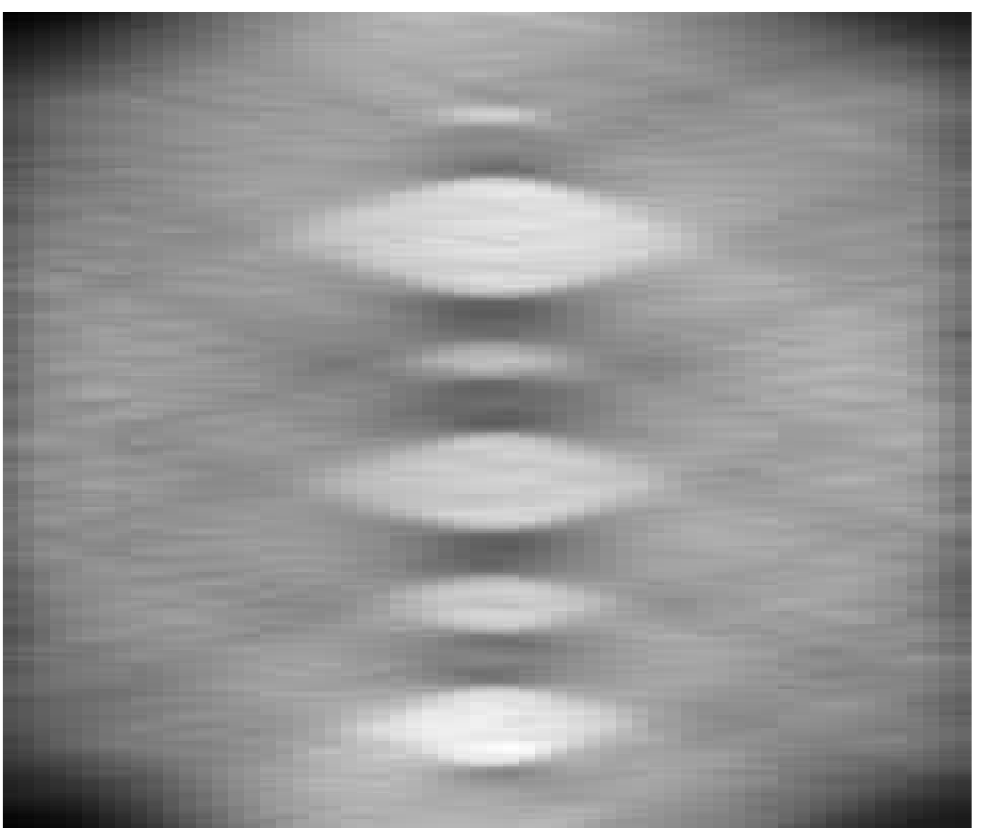}
\includegraphics[width=1.1in]{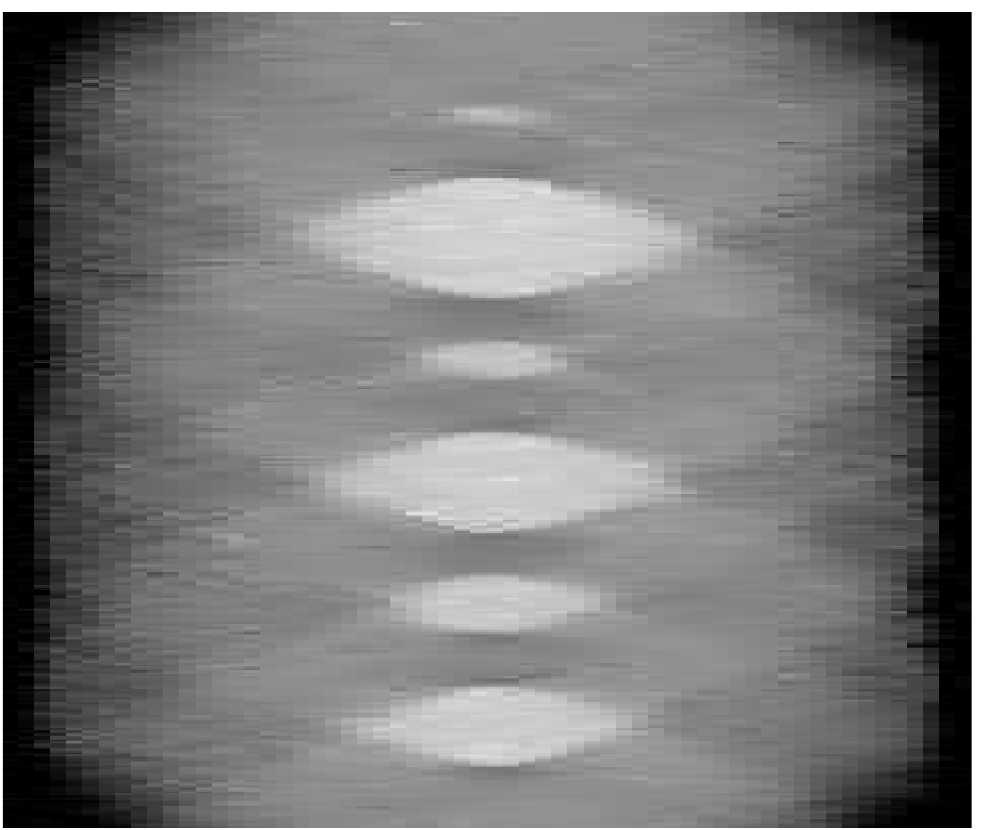}\\
\includegraphics[width=1.1in]{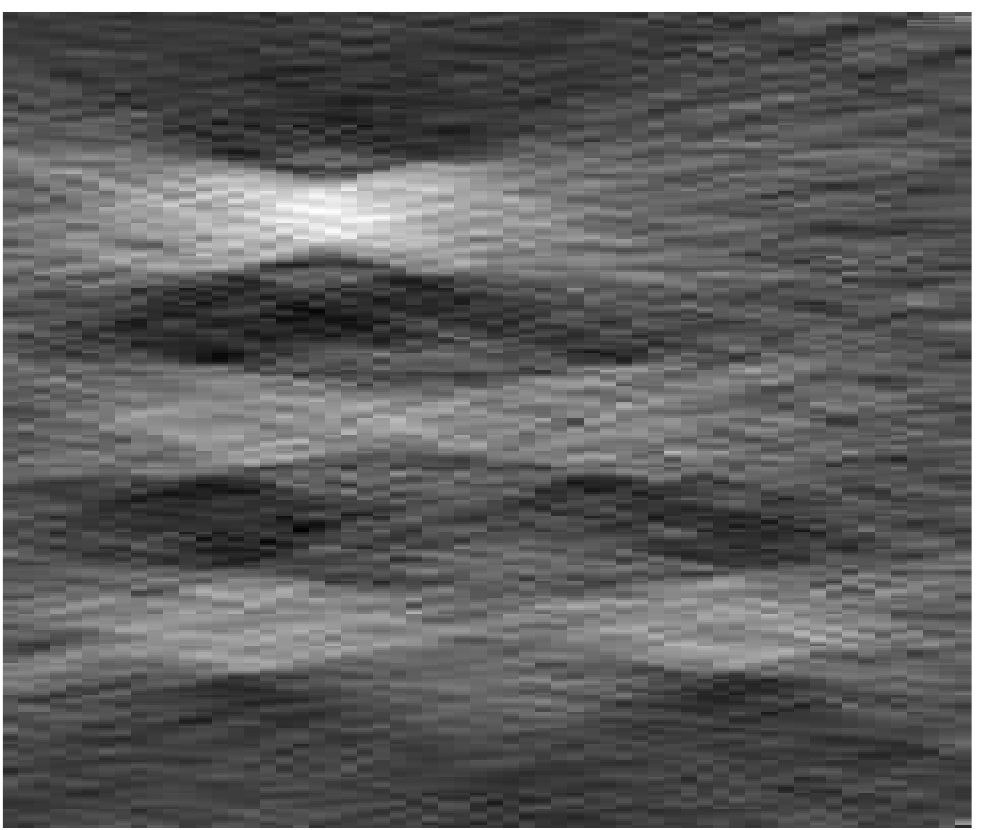}
\includegraphics[width=1.1in]{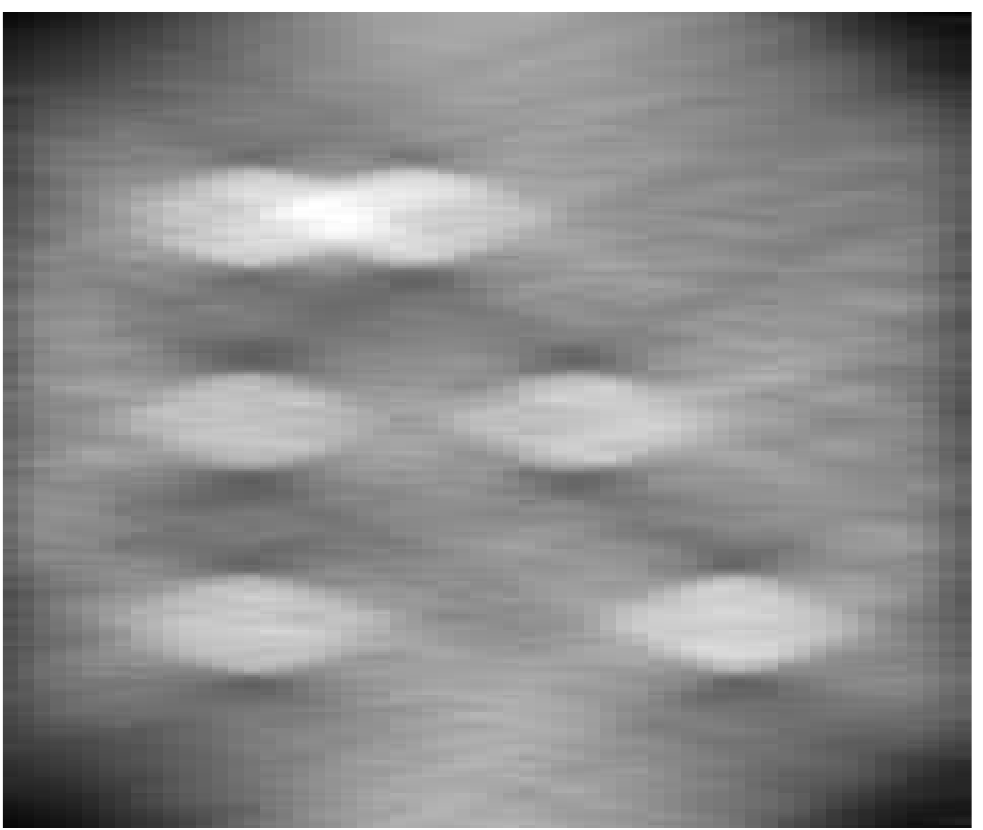}
\includegraphics[width=1.1in]{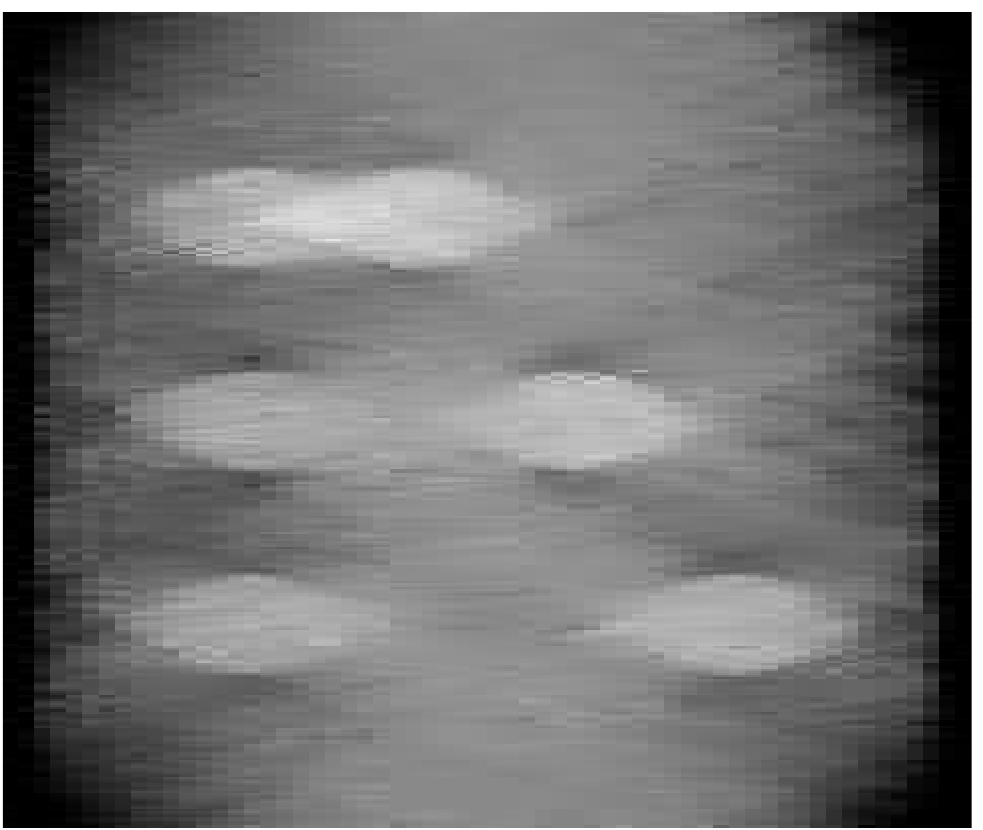}
\caption{
2D slices within images reconstructed by use of 
the FBP (left column), the EM (middle column), and the TV-minization
(right column) algorithms, respectively. The corresponding 
true images are displayed in 
Figs.~\ref{fig:CUTXY}b, \ref{fig:CUTXY}c and \ref{fig:CUTXY}d.
}
\label{fig:CUTdepth}
\end{figure}

We also show in Fig.~\ref{fig:CUTdepth} 
the images from noisy data within planes
that are parallel to the source-motion plane
and perpendicular to the detector. 
The true images corresponding to those
in rows 1, 2, and 3 of Fig.~\ref{fig:CUTdepth}
are shown in in Figs.~\ref{fig:phantom}(b),
\ref{fig:phantom}(c), and \ref{fig:phantom}(d),
respectively. As the results indicate,
none of the algorithms recover the true shape of the
the ellipsoids of varying flatness in the first row. 
Rather, the variation in flatness results in a
variation of contrast. Furthermore, an increase in overall intensity
across the uniform pectoralis muscle can be observed for the iterative
algorithms.  For all reconstructions, as shown in
the second row of Fig.~\ref{fig:CUTdepth}, 
the spheres of different sizes are elongated
due to the limited angular range. Objects are
most uniform in the TV reconstruction and most blurred in the FBP
slice. The improved in-depth resolution of the iterative algorithms is
also demonstrated for stacked spheres shown
in the third row of Fig.~\ref{fig:CUTdepth}. 
Both algorithms clearly separate spheres spaced by 2d, 
while FBP still shows some overlap. 
In all slices obtained through FBP reconstruction, quantum noise is 
quite pronounced. For the iterative algorithms, structure noise is
dominant.

\begin{figure}
\centering
\includegraphics[width=1.0in]{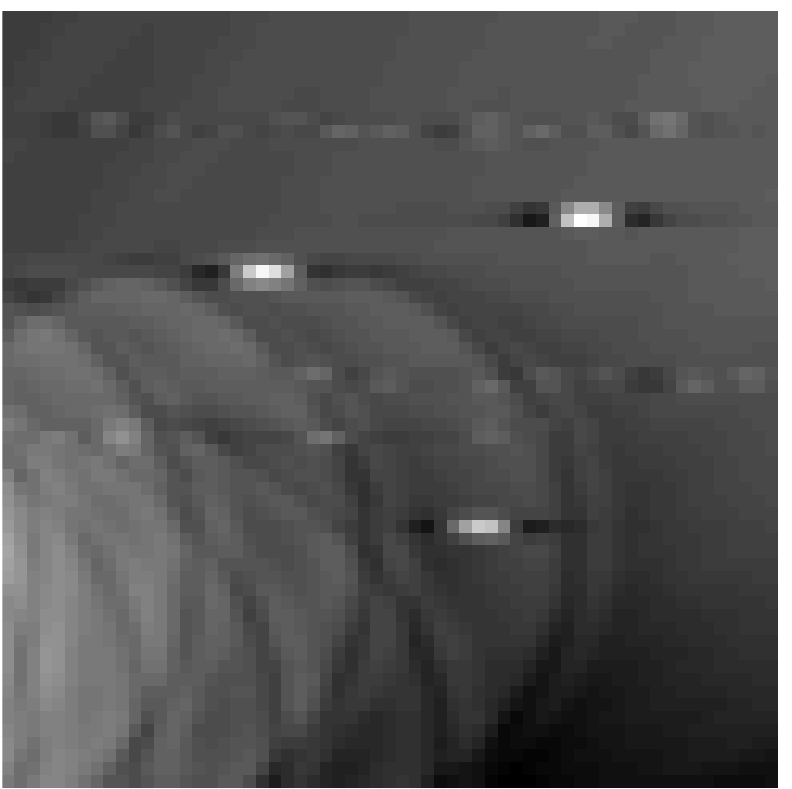}
\includegraphics[width=1.0in]{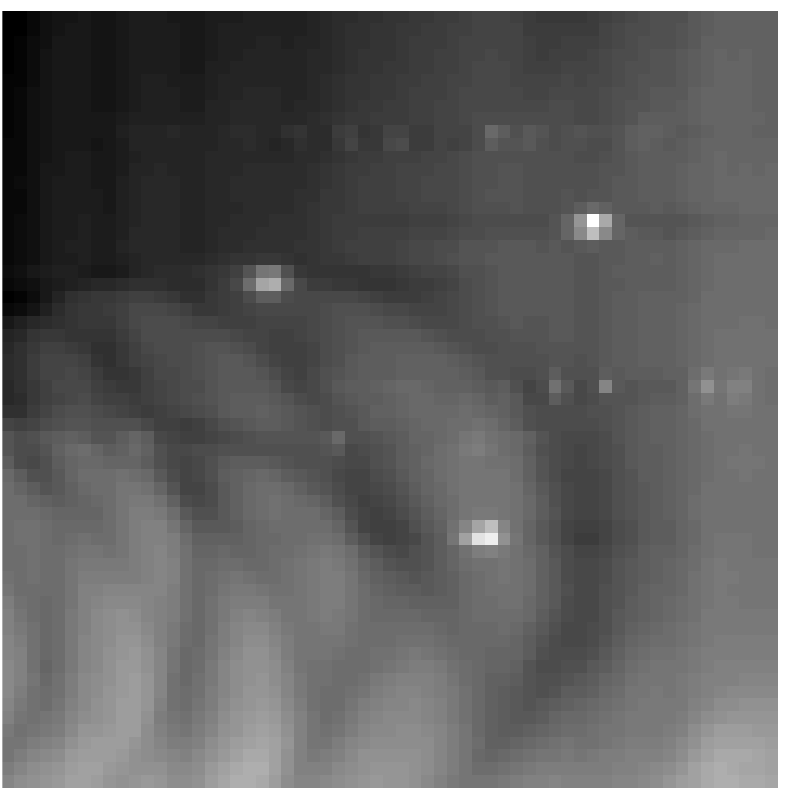}
\includegraphics[width=1.0in]{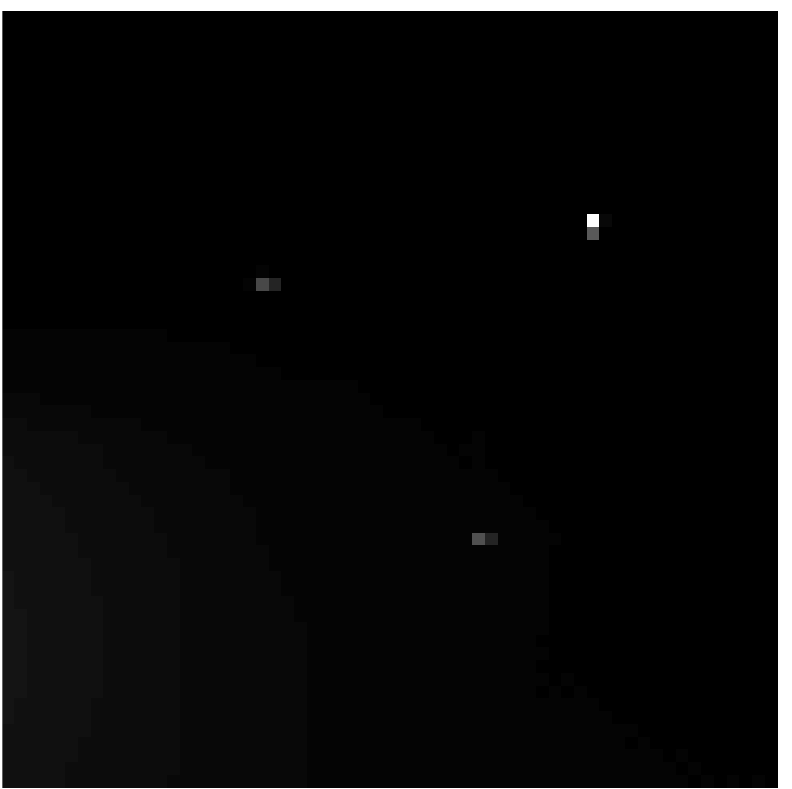}\\
\includegraphics[width=1.0in]{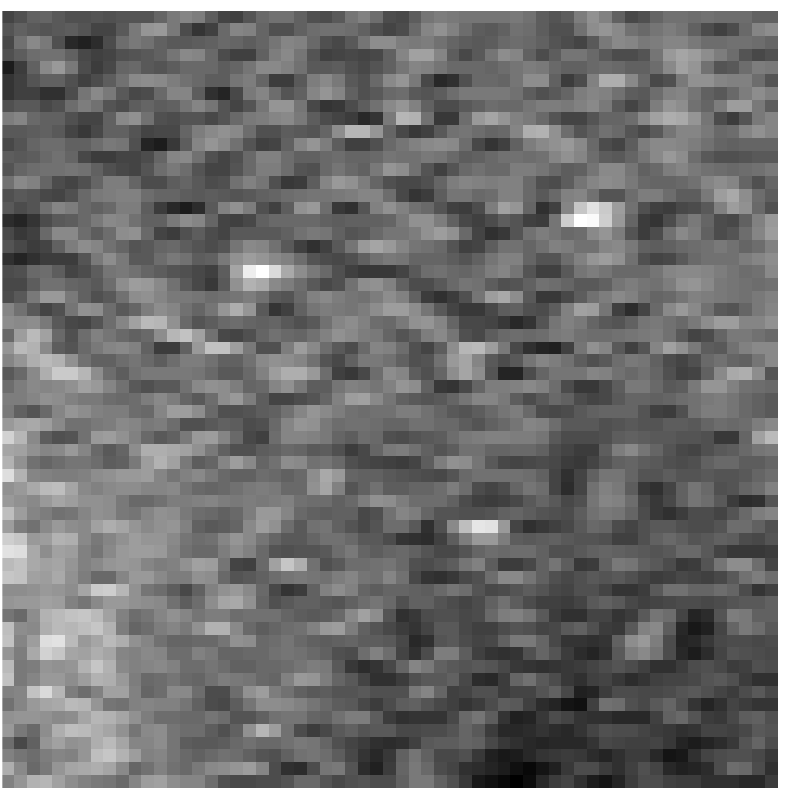}
\includegraphics[width=1.0in]{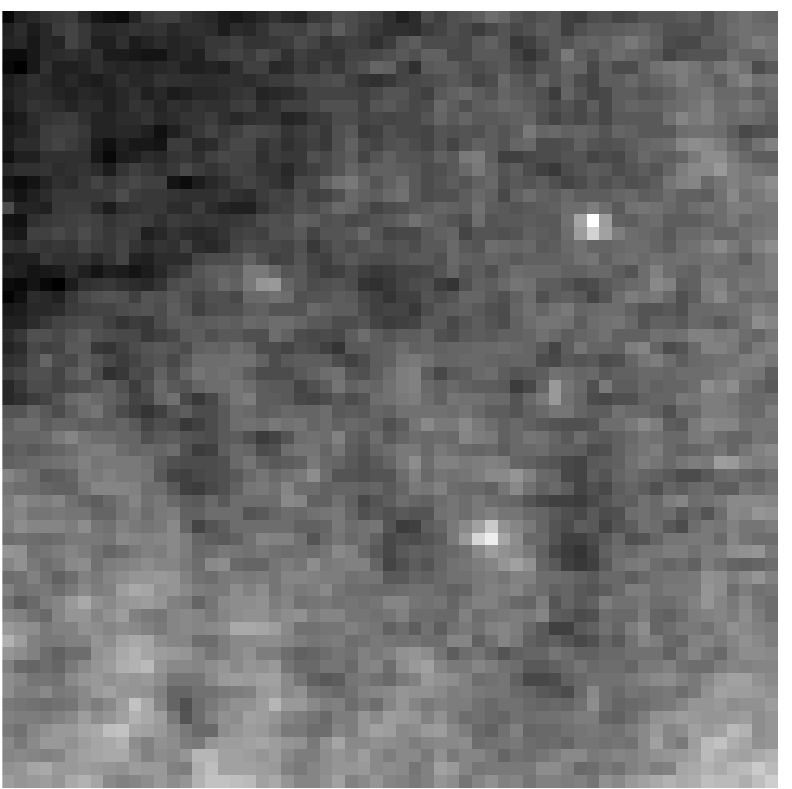}
\includegraphics[width=1.0in]{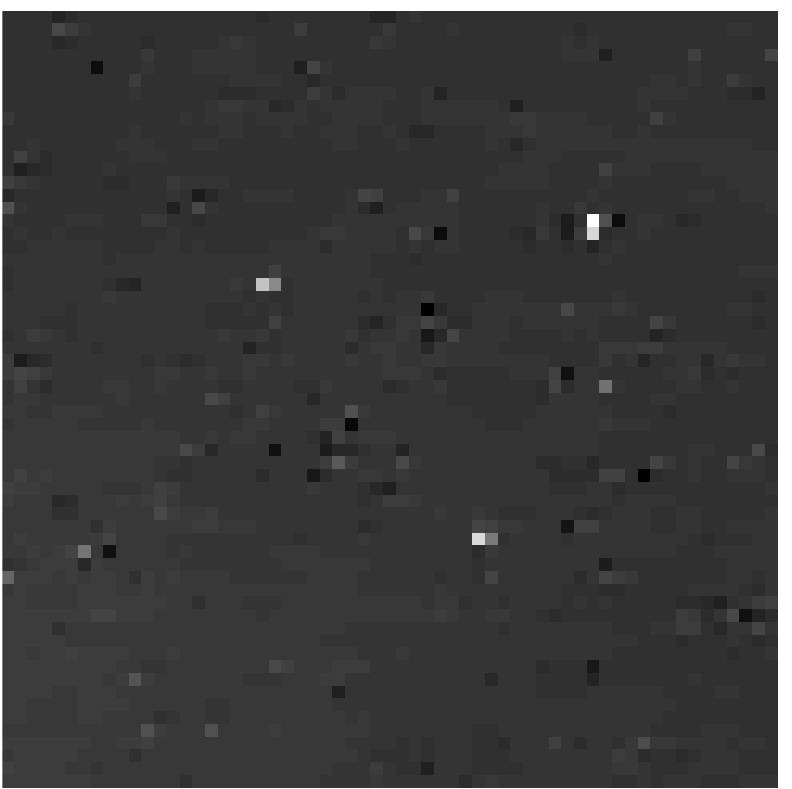}
\caption{Images of a set of small microcalcifications ($0.15$~mm in diameter)
within the 2D in-focus plane reconstructed
by use of the FBP (left column), EM (middle column), 
and, TV-minimization (right column), respectively,
from noiseless (row 1) and noisy (row 2) data.} 
\label{fig:SmallCalcs}
\end{figure}

Finally, we show in Fig.~\ref{fig:SmallCalcs} a region-of-interest
(ROI) in the in-focus plane of a microcalcification (MC) cluster. 
The top row shows reconstructions from
noiseless data, the bottom row shows reconstructions from noisy
data. The entire cluster consists of 6 MC, with
another triangular arrangement 6~mm below the one depicted. In the
noiseless images, patterns caused by streaking artifacts can be
perceived for the FBP and EM reconstructions. 
Image quality in the TV reconstruction is clearly superior in this
situation. 

\section{Conclusion}
We presented a systematic comparison of three reconstruction methods
(FBP, EM and TV), for noiseless and noisy phantom data acquired for a
tomosynthesis scanning configuration. The comparison was based on a
phantom that was specifically designed for evaluating tomosynthesis
images. Our study demonstrates that all reconstructed images exhibit
artifacts, because of the incomplete projection data available for
reconstruction. However, in-depth resolution was higher in images from
iterative reconstructions, and therefore ghosting artifacts were also
less conspicuous in the iterative reconstructions. Furthermore, structure
noise from out-of-plane objects was dominant for iterative
reconstructions, while quantum noise became visible in the FBP
reconstruction. The FBP reconstruction also showed artifacts caused by
the ramp filtering. In summary, our study indicates that iterative
algorithms may provide higher image quality for tomosynthesis image
reconstruction than FBP-based reconstruction. 

\section*{Acknowledgment}
This work was funded in part by NIH grants R33 CA109963, R01 EB000225, and
K01 EB003913 (EYS).

\bibliography{tomo,sdt,imaging,CT,detector}

\begin{thebibliography}{10}

\bibitem{CancerFacts:2006}
{A}merican~{C}ancer {S}ociety.
\newblock {\em {C}ancer {F}acts and {F}igures 2006}.
\newblock {A}merican {C}ancer {S}ociety, Atlanta, 2006.

\bibitem{Burgess:2001}
A~E Burgess, F~L Jacobson, and P~F Judy.
\newblock Human observer detection experiments with mammograms and power-law
  noise.
\newblock {\em Med. Phys.}, 28:419--437, 2001.

\bibitem{Dobbins:2003}
J~T Dobbins, III and D~J Godfrey.
\newblock Digital x-ray tomosynthesis: current state of the art and clinical
  potential.
\newblock {\em Phys. Med. Biol.}, 48:R65--R106, 2003.

\bibitem{Niklason:1997}
L~T Niklason, B~T Christian, L~E Niklason, D~B Kopans, D~E Castleberry, B~H
  Opsahl-{O}ng, C~E Landberg, P~J Slanetz, A~A Giardino, R~Moore, D~Albagli,
  M~C De{J}ule, P~F Fitzgerald, D~F Fobare, B~W Giambattista, R~F Kwasnick,
  J~Liu, S~J Lubowski, G~E Possin, J~F Richotte, C-Y Wei, and R~F Wirth.
\newblock Digital tomosynthesis in breast imaging.
\newblock {\em Radiology}, 205:399--406, 1997.

\bibitem{Ren:2005}
B~Ren, C~Ruth, J~Stein, A~Smith, I~Shaw, and Z~Jing.
\newblock Design and performance of the prototype full field breast
  tomosynthesis system with selenium-based flat panel detector.
\newblock In {\em Proc. SPIE}, volume 5745, page 550, 2005.

\bibitem{Mertelmeier:2006}
T~Mertelmeier, J~Orman, W~Haerer, and M~K Dudam.
\newblock Optimizing filtered backprojection reconstruction for a breast
  tomosynthesis prototype device.
\newblock In {\em Proc. SPIE}, volume 6142, page 61420F, 2006.

\bibitem{Maidment:2006}
A~D~A Maidment, C~Ullberg, K~Lindman, L~Adel{\"{o}}w, J~Egerstr{\"{o}}m,
  M~Eklund, T~Francke, U~Jordung, T~Kristoffersson, L~Lindqvist, D~Marchal,
  H~Olla, E~Penton, J~Rantanen, S~Solokov, N~Weber, and H~Westerberg.
\newblock Evaluation of a photon-counting breast tomosynthesis imaging system.
\newblock In {\em Proc. SPIE}, volume 6142, page 61420B, 2006.

\bibitem{Zhang:2006}
Y~H Zhang, H~P Chan, B~Sahiner, J~Wei, M~M Goodsitt, L~M Hadjiiski, J~Ge, and
  C~A Zhou.
\newblock A comparative study of limited-angle cone-beam reconstruction methods
  for breast tomosynthesis.
\newblock {\em Med. Phys.}, 33:3781--3795, 2006.

\bibitem{Wu:2003}
T~Wu, A~Stewart, M~Stanton, T~Mc{C}auley, W~Phillips, D~B Kopans, R~H Moore,
  J~W Eberhard, B~Opsahl-{O}ng, L~Niklason, and M~B Williams.
\newblock Tomographic mammography using a limited number of low-dose cone-beam
  projection images.
\newblock {\em Med. Phys.}, 30:365--380, 2003.

\bibitem{Johns:1987}
P~C Johns and M~J Yaffe.
\newblock X-ray characterisation of normal and neoplastic breast tissues.
\newblock {\em Phys. Med. Biol.}, 32:675--695, 1987.

\bibitem{Wu:2004}
T~Wu, R~H Moore, E~A Rafferty, and D~B Kopans.
\newblock A comparison of reconstruction algorithms for breast tomosynthesis.
\newblock {\em Med. Phys.}, 31:2636--2647, 2004.

\bibitem{Sidky:2006}
EY~Sidky, CM~Kao, and X~Pan.
\newblock Accurate image reconstruction from few-views and limited-angle data
  in divergent-beam {CT}.
\newblock {\em Journal of X-Ray Science and Technology}, 14:119--139, 2006.

\end{thebibliography}
\bibliographystyle{unsrt}

\end{document}